\documentclass[conf]{new-aiaa} 
\usepackage[utf8]{inputenc}
\usepackage{textcomp}
\usepackage{multicol}
\usepackage{lscape}

\usepackage{cleveref}
\usepackage{algorithm}
\usepackage{algpseudocode}
\usepackage{graphicx}
\usepackage{amsmath}
\usepackage[version=4]{mhchem}
\usepackage{siunitx}
\usepackage{longtable,tabularx}
\usepackage{xcolor} 
\usepackage{wrapfig} 
\usepackage{subcaption} 
\usepackage{array}
\newcolumntype{L}[1]{>{\raggedright\let\newline\\\arraybackslash\hspace{0pt}}m{#1}}
\newcolumntype{C}[1]{>{\centering\let\newline\\\arraybackslash\hspace{0pt}}m{#1}}
\newcolumntype{R}[1]{>{\raggedleft\let\newline\\\arraybackslash\hspace{0pt}}m{#1}}

\usepackage{tikz}
\usetikzlibrary{shapes}

\newcommand{\rev}[1]{\textcolor{black}{#1}}

\usepackage{nomencl}
\makenomenclature

\title{Safety, Trust, and Ethics Considerations for \\Human-AI Teaming in Aerospace Control} 

\author{Kerianne L. Hobbs \footnote{Approved for Public Release. Case Numbers 2023-0428, AFRL-2023-1934.}\footnote{Safe Autonomy Lead, Autonomy Capability Team 3, 2241 Avionics Circle, AIAA Associate Fellow}}
\affil{Air Force Research Laboratory, Wright-Patterson AFB, OH, 45433}
\author{Bernard Li\footnote{Bluehalo LLC, Air Force Research Laboratory, 3550 Aberdeen Avenue Southeast, Building 577, Kirtland Air Force Base, New Mexico, 87117} }
\affil{Air Force Research Laboratory, Kirtland AFB, New Mexico, 87117}

\begin{document}

\maketitle

\begin{abstract}
Designing a safe, trusted, and ethical AI may be practically impossible; however, designing AI with safe, trusted, and ethical use in mind is possible and necessary in safety and mission-critical domains like aerospace. Safe, trusted, and ethical use of AI are often used interchangeably; however, a system can be safely used but not trusted or ethical, have a trusted use that is not safe or ethical, and have an ethical use that is not safe or trusted. This manuscript serves as a primer to illuminate the nuanced differences between these concepts, with a specific focus on applications of Human-AI teaming in aerospace system control, where humans may be in, on, or out-of-the-loop of decision-making. 
\end{abstract}




\section{Introduction}

\lettrine{O}{n} May 11th, 1997, Garry Kasparov became the first World Chess Champion to lose to a computer, IBM’s DeepBlue. A year later, Kasparov created a modified version of chess called “advanced chess,” where instead of a human playing against a human, a human-computer team would play against another human-computer team \cite{kasparovarticle}. In the first advanced chess tournament hosted in 2005, the person on the winning team was neither a grandmaster nor was the algorithm on the winning team the most advanced. Instead, an amateur who was more trusting of their algorithms to make the right decisions due to his inexperience, while also training with the algorithm to recognize its flaws and weaknesses, took home the trophy with 3 relatively weak laptops that he could manage, each giving separate recommendations \cite{kasparovarticle}. Kasparov, and many others, realized that characteristics such as trust played a large role in human-algorithm team performance. Pairing humans with increasingly intelligent algorithms may result in significant performance gains over the next few years, especially highly capable deep learning models. 

\textit{Artificial intelligence} (AI), is defined as the ability of a computer or robot to perform tasks that normally require human intelligence \cite{AI2023}, and \textit{machine learning} (ML), is defined as the use and development of computer systems that are able to learn and adapt without following explicit instructions, by using algorithms and statistical models to analyze and draw inferences from patterns in data \cite{ML2023}. In ML, a \textit{model} is a fixed structure (such as a deep neural network) with modifiable parameters that encodes an input-output relationship \cite{alpaydin2016machine}. Within ML, there are three broad categories: supervised, unsupervised, and reinforcement learning. In \textit{supervised learning}, labeled input/output data is used to train a model to generate the correct output for any input \cite{alpaydin2016machine}. \textit{Unsupervised learning} uses unlabeled data to find trends or clusters of output data. In \textit{reinforcement learning (RL)}, “learning to take actions,” the model is referred to as an \textit{agent} that learns a sequence of actions (outputs) that returns a maximum reward. RL is based on the concept of operant conditioning where desirable behavior is reinforced with rewards (positive reinforcement) and bad behavior is penalized (negative reinforcement), and the agent learns which actions to take to maximize reward through trial and error \cite{sutton2018reinforcement}. Applied to aerospace or robotics, AI and ML applications (in particular, RL) are often embodied as an \textit{autonomous system}, which is a cyber-physical system that is given delegated and bounded authority by a human to make decisions and take actions in the world. 

While there are cases in which a model can be trained using supervised learning from a look-up table of optimal data points, such as the neural network compression of ACAS-XU \cite{julian2019deep}, RL approaches are attractive, in part due to similarities to control theory. RL and optimal control share the concept of a system \textit{state} (variables that describe the plant / \textit{RL environment}, such as position and velocity), a notion of \textit{partial observability} (where a measurement / \textit{RL observation} can be used to estimate the state), a discrete or continuous control input / \textit{RL action} is determined by a control law / \textit{RL policy} (a function of the current state or partial observation that is optimized to minimize a cost function in controls or maximize a \textit{reward function} in RL). These similarities are summarized in Table \ref{ControlsvsRL}. 
\begin{table} [htb!]
\caption{Comparison of Control Theory and Reinforcement Learning Concepts \cite{ravaioli2022safe}}
\label{ControlsvsRL}
\centering
\begin{tabular}{|c | c|}
\hline
{\bfseries Control Theory}   & {\bfseries Reinforcement Learning}  \\ \hline
{plant}   & {environment}  \\
state &  state  \\ 
measurement &   observation  \\  
{controller}   & {agent}  \\
{control law}   & {policy}  \\
control input & action\\ 
cost function & reward function\\ 
\hline
\end{tabular}
\end{table}
In aviation, RL has been used to develop autonomous drone racing solutions that outperformed human experts and optimal control strategies \cite{song2023reaching}, and has the potential to improve air-based wildfire monitoring \cite{julian2019distributed}, urban air mobility traffic deconfliction \cite{deniz2022multi}, air traffic flow \cite{crespo2012reinforcement}, landings \cite{tang2020deep},  maintenance scheduling \cite{andrade2021aircraft}, acrobatic maneuvering \cite{clarke2020deep}, control system designs \cite{konatala2021reinforcement,waslander2005multi,bohn2019deep,tandale2004preliminary,de2019reinforcement,woodbury2014autonomous}, terminal area operations and planning \cite{balakrishna2010accuracy}, military engagement scenarios such as the AlphaDogfight~\cite{pope2021hierarchical}, and other areas. In the spacecraft domain, RL has been applied to research models for tasks such as hovering near asteroids \cite{gaudet2012robust}, ground observation missions \cite{harris2020spacecraft}, proximity operations and docking \cite{scorsoglio2019actor,broida2019spacecraft,dunlap2023run}, autonomous spacecraft maneuvering \cite{oestreich2020autonomous,hovell2020deep}, autonomous mapping of asteroids \cite{chan2019autonomous}, planetary landing \cite{gaudet2020deep,gaudet2020adaptive}, and other applications.

Responsible AI development is of high concern in the United States as evidenced by the recent "Executive Order on the Safe, Secure, and Trustworthy Development and Use of Artificial Intelligence" \cite{Biden2023Executive}. Safe, trusted, and ethical use of AI are of particular concern in Aerospace because contrary to many applications of AI and ML, aerospace systems feature high safety and reliability requirements. For example, reliability requirements are on the order of 10$^{-9}$ catastrophic fault rate for civil aircraft avionics \cite{rushby2009software}. The authors argue that anthropomorphizing AI to have qualities such as ethical behavior, places unrealistic expectations on AI. Designing a safe, trusted, or ethical AI system may be practically impossible. However, safe, trusted, and ethical \textit{use} of an AI system from ideation, through development and operational use is feasible. Safety, trust and ethics are related emergent properties of human-AI teams. \textit{Safety} is freedom from harm during operations, \textit{trust} is a willingness to accept vulnerability in situations characterized by uncertainty \cite{lee2004trust}, \textit{ethics} in this context are rules created by societies and cultures governing moral and just usage of technology, and \textit{emergent properties} are characteristics that an entity gains when it becomes part of a larger system. This publication explores safety, trust, and ethics context that should be taken into consideration for an AI that makes decisions and controls behavior of an aerospace system in a team with a human.
\section{Humans In, On, and Out-of-the-Loop}
\begin{figure} [htb]
\centering
\begin{subfigure}{.4\textwidth}
  \centering
  \includegraphics[width=.94\linewidth]{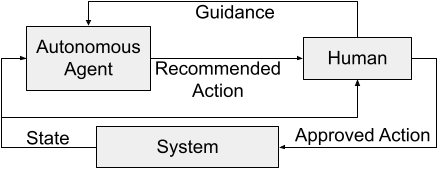}
  \caption{Human-in-the-loop}
  \label{fig:sub1}
\end{subfigure}%
\begin{subfigure}{.17\textwidth}
  \centering
  \includegraphics[width=.9\linewidth]{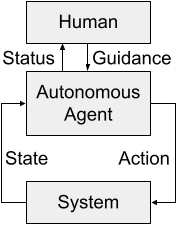}
  \caption{Human-on-the-loop}
  \label{fig:sub2}
\end{subfigure}
\begin{subfigure}{.17\textwidth}
  \centering
  \includegraphics[width=.9\linewidth]{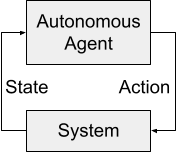}
  \caption{Human-out-of-the-loop}
  \label{fig:sub3}
\end{subfigure}
\caption{Comparison of human-AI teaming for control in which the human is in, on, or out-of-the-loop.}
\label{fig:loops}
\end{figure}

Control systems and AI models that take actions based on observations of the state, make decisions in a loop. The relationship between humans and that decision loop, is often described as in, on, or out-of-the-loop, and is depicted in Figure \ref{fig:loops}. For \textit{human-in-the-loop} systems, no action is taken without the approval of a human. Actions are taken at a slow enough rate that both the human and the autonomous agent know the state of the system, a human provides guidance to an autonomous agent, the human has time to evaluate the recommended action from the autonomous agent, and the human approves the action that is sent to the system to execute. Human-in-the-loop teams are more appropriate for discrete actions where long periods of time are available between decisions, such as planning a trajectory. \textit{Human-on-the-loop} systems feature a human that gets status updates from the autonomous agent and the state of the state of the system under control, and provides guidance to the autonomous agent as it takes independent actions. This category is appropriate in situations in which human decision making may not be fast enough, human approval of every decision is not required, and a human supervisory role makes more sense. \textit{Human-out-of-the-loop} teams feature an autonomous agent that observes the state and takes actions, perhaps with some initial guidance from a human operator, but otherwise independently. This approach may be appropriate in situations where regular communication with a human operator is not possible.  More research is needed to explore differences in the safety, trust, and ethical principles applied to designing models that make decisions and take actions at different levels of human involvement. The rest of this manuscript considers decisions takes a very general approach to safety, trust, and ethics. 

\section{Safety}

Although trusted and ethical use of AI in a human-AI team may be difficult to define formally, safety has been a huge part of aerospace for more than a century, and can be formally evaluated with respect to certification criteria, risk definitions, and design principles. The Center for Security and Emerging Technology defines AI safety as "an area of machine learning research that aims to identify causes of unintended behavior in machine learning systems and develop tools to ensure these systems work safely and reliably," and defines safety AI issues in categories of specification, assurance, and robustness \cite{rudner2021key}. Specification describes the challenge of describing the intended system behavior, sometimes in the form of a reward function during training, sometimes in the form of a mathematically or logically defined formal specification \cite{hobbs2021formal}, and the opportunity for misspecification that leads to unintended behavior. Assurance is analogous to interpretability of the AI in the definition, and robustness refers to the ability of the AI to behave reliably in a wide range of decisions. While these issues are important to human-AI teams in aerospace, they are not sufficient. This section describes several other important considerations for human-AI teams in aerospace control.


\subsection{Standards for Aerospace Safety}

Several standards provide guidance for assuring safety in the design of aerospace systems, including Aerospace Recommended Practice (ARP) from the Society of Automotive Engineers (SAE) International, Radio Technical Commission for Aeronautics (RTCA) guidance, American Society for Testing and Materials (ASTM) standards, as well as military and federal standards, handbooks, and regulations. Where appropriate, the following standards, handbooks, and guidance should be consulted:
\begin{itemize}
\item ARP4761: Guidelines and Methods for Conducting the Safety Assessment Process on Civil Airborne Systems and Equipment \cite{ARP4761}
\item ARP4754: Guidelines For Development Of Civil Aircraft and Systems \cite{ARP4754A}
\item RTCA DO-254: Design Assurance Guidance for Airborne Electronic Hardware \cite{DO-254}
\item RTCA DO-178: Software Considerations in Airborne Systems and Equipment Certification \cite{DO-178C}
\item RTCA DO-333: Formal Methods Supplement to DO-178C and DO-278A \cite{DO-333}
\item ASTM3269-21 (update from ASTM 3269-17): Standard Practice for Methods to Safely Bound Flight Behavior of Unmanned Aircraft Systems Containing Complex Functions \cite{ASTM3269-17, ASTM3269-21}
\item MIL-HDBK-516C: Airworthiness Certification Criteria \cite{MILHDBK516C}
\item MIL-STD-882E: System Safety \cite{MILSTD882E}
\item FAA's Standard Airworthiness Certification Regulations \cite{FAAAirworthiness}
\end{itemize}

In aerospace, roadmaps such as Concepts of Design Assurance for Neural Networks (CoDANN) I and II \cite{CoDANNI,CoDANNII} have primarily focused on supervised and unsupervised learning applications such as machine perception. More research is needed to develop criteria for RL-based autonomy solutions.

\subsection{Safety by Design}
 
\begin{figure}[htb!]
    \centering
    \includegraphics[width = .49\textwidth]{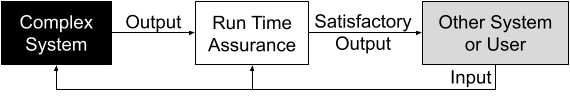}
    \caption{A generalization of a Run Time Assurance (RTA) architecture }
    \label{fig:GeneralRTA}
\end{figure}

\textit{Run time assurance} systems monitor the output of an advanced, primary controller and intervene when necessary to assure safety \cite{hobbs2023run}, as depicted in Figure \ref{fig:GeneralRTA}. Until technology advances enough to prove properties of neural network control systems, run time assurance is the most likely near term path to certify advanced control systems. The ASTM3269-21 \cite{ASTM3269-21}, and its predecessor, are the first examples of run time assurance certification recommendations.

A generalization of a Run Time Assurance (RTA) architecture is shown in \ref{fig:GeneralRTA}. In the figure, a black-box complex function could be a human controller, a trained neural network, or some other complex function that is difficult to verify. This black box controller produces an output, such as a thrust command, based on an input, such as an observation of position and velocity. A white-box RTA system monitors the black box output, and if necessary, modifies the output to assure adherence to safety constraints. Here, a \textit{black-box} system can only be viewed in terms of inputs and outputs while the internal mechanisms are opaque, \textit{white-box} system inner workings are fully known, and \textit{gray-boxes} internal mechanisms are partially known. Note that while the weights, biases, activation functions, connections, and other features of neural networks can be known, it is generally argued that their combination and interactions are so complex as to effectively be a black box to a human user.

The Automatic Ground Collision Avoidance System (Auto GCAS) and the Automatic Air Collision Avoidance System (Auto ACAS) \cite{burns2011auto, turner2012automatic, swihart2011design, swihart2011automatic, lyons2016trust, wadley2013development, sorokowski2015small} are interesting case studies here in two separate contexts. First, both are examples of an aerospace run time assurance system, and second, both are arguably examples of the successful use of human-autonomy teaming; however, there is no machine learning or neural network used in Auto GCAS or Auto ACAS. In Auto GCAS and Auto ACAS, a human provides the primary performance control to complete a mission (black-box complex function), while a white-box autonomous system monitors the state of the aircraft relative to the ground or other aircraft, and intervenes when necessary to avoid a collision. Auto GCAS was transitioned to the F-16 fleet with digital flight control computers in fall of 2014 and has since been credited with saving the lives of 13 pilots in 12 F-16s. It is also a unique aerospace system that takes control from a pilot in life or death situations, is the result of more than three decades of research, and has been studied extensively by psychologists as a human-automation teaming example \cite{ho2017trust,lyons2016trust}.

While there is debate whether trust \cite{idapaper} and ethics \cite{ethicsHAIT} can be built into a system, safety is agreed upon to be a necessary design criteria of any autonomous robotic system interfacing in real-world scenarios, especially in proximity to humans. Auto GCAS \cite{hobbs2020elicitation} which has several critical design principles that contribute to safety, as follows: 

\begin{itemize}
    \item \textit{understandable response}: the human teammate should have an acceptable understanding of what the autonomous agent is doing, e.g. in Auto GCAS the response is a simple roll to wings level and 5g pull maneuver, 
    \item \textit{appropriate transparency}: visual or auditory explanations of the AI system perceptions should be balanced, so they do not result in overtrust of the system or create a distraction or nuisance to the operator, 
    \item \textit{nuisance freedom}: the responses of the AI agent should not be so excessive that they compromise performance of the team, e.g. in Auto GCAS, the system would only maneuver at a point in time beyond when a human pilot would maneuver to reduce possible mission impacts, 
    \item \textit{high level requirements precedence}: the human-AI team should agree on the order of high level requirements, e.g. on Auto GCAS the requirements were in the order of "do no harm, do not interfere, prevent collisions," in which the primary task of the "AI" was placed at a lower priority than not interfering with the mission or causing harm to the human or system under control,
    \item \textit{system-wide integrity management}: a significant element of ensuring system safety is recognizing and accommodating failures. Most aircraft are triple or quad redundant and use voting schemes to identify failures. That said, AI, depending on its form, can be brittle and fail catastrophically in non-traditional ways, so redundancy may not be the best antidote. Auto GCAS uses a hardware and software checks to ensure safety of maneuver requests, 
    \item \textit{variable risk tolerance}: different missions may have different definitions of what constitutes "nuisance" AI, so an ability to modify parameters, such as those related to risk are necessary.
\end{itemize}

\subsection{Risk}
\begin{figure}[htb!]
\centering
\begin{subfigure}{.59\textwidth}
   \centering
    \includegraphics[width = .95\textwidth]{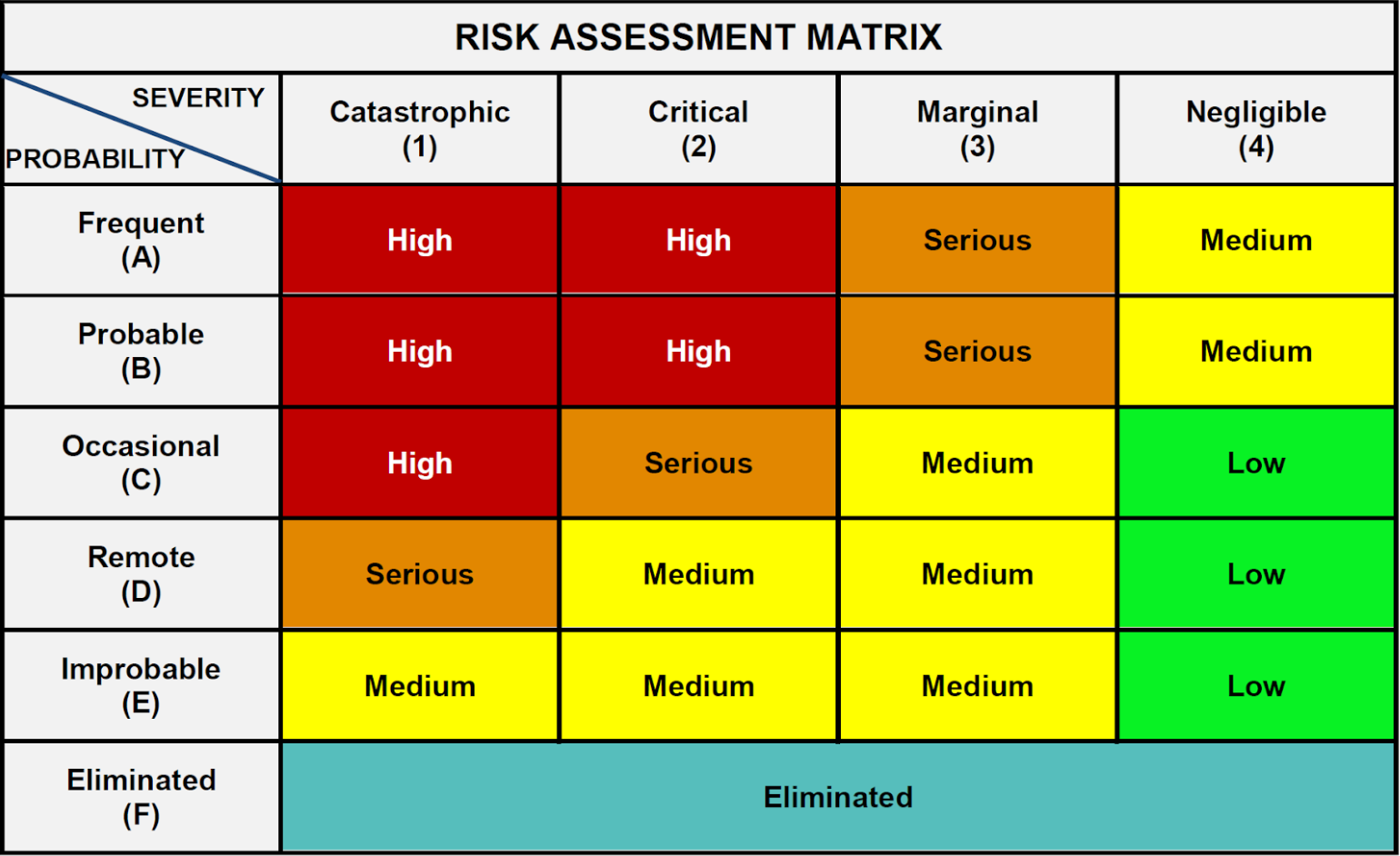}
    \caption{Military Risk Assessment Matrix \cite{MILSTD882E}}
    \label{fig:MIL-STD-882E}
\end{subfigure}%
\begin{subfigure}{.4\textwidth}
   \centering
    \includegraphics[width = .95\textwidth]{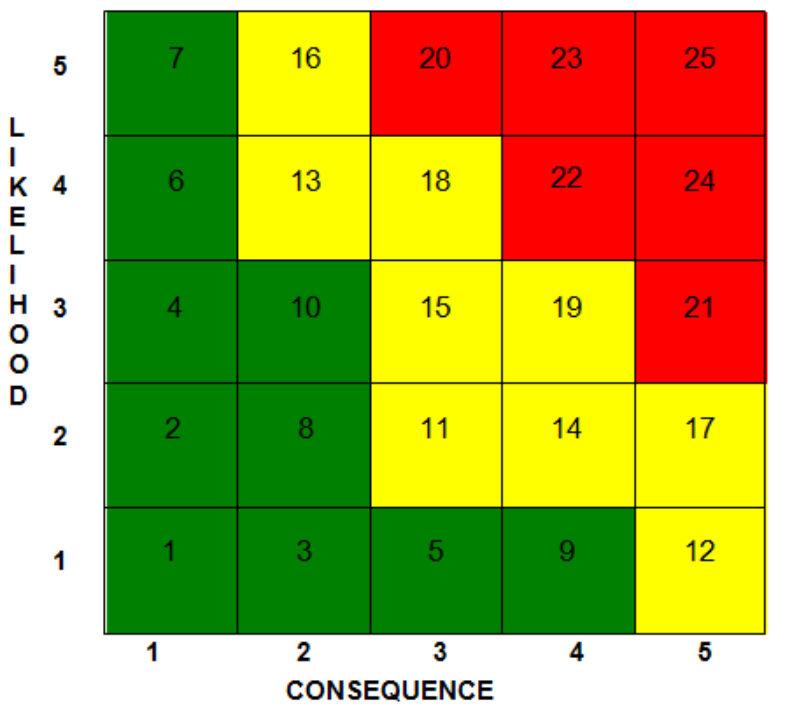}
    \caption{NASA Risk Assessment Matrix \cite{S3001}}
    \label{fig:S3001Risk Matrix}
\end{subfigure}
\caption{NASA and Military risk assessment matrices.}
\label{fig:Riskassessments}
\end{figure}

The working definition of trust includes a willingness to be vulnerable, which is related to safety and risk tolerance. Risk is defined by government organizations like the military and NASA in terms of likelihood and severity of particular risks. Definitions of probability and severity in these contexts are provided in Tables \ref{tab:probability}-\ref{tab:severity}. These probabilities and severities are then used to assign a risk level via a risk matrix such as Figure \ref{fig:MIL-STD-882E} for the military and Figure \ref{fig:S3001Risk Matrix} for NASA. Also of relevance to AI and autonomy software is the U.S. Military definition of software safety criticality levels: no safety impact, influential, redundant fault-tolerant, semi-autonomous, and autonomous control.

\begin{table*}[htb!]
    \centering
    \caption{Probability Categories from MIL-STD-882E \cite{MILSTD882E} and NASA S3001 \cite{S3001}}
    \begin{tabular}{|p{1cm}|p{7cm}|p{1cm}|p{2.6cm}|p{2.5cm}|}\hline
        \raggedright 882E Level  & 882E Criteria & \raggedright S3001 Level & S3001 Criteria & S3001 Probability \\\hline
        A & Likely to occur often in the life of an item. &5 & Near certainty &$p>80$\%\\\hline
        B & Will occur several times in the life of an item.&4 & Highly Likely &80\%$>p>60$\%\\\hline
        C & Likely to occur sometime in the life of an item. &3& Likely  &60\%$>p>40$\% \\\hline
        D & Unlikely, but possible to occur in the life of an item. & 2 & Low likelihood &40\%$>p>20$\%\\\hline
        E & So unlikely, it can be assumed occurrence may not be experienced in the life of an item. &1 &Not likely &20\%$>p>0$\%\\\hline
        F & Incapable of occurrence. This level is used when potential hazards are identified and later eliminated.& & & \\\hline
    \end{tabular}
    \label{tab:probability}
\end{table*}
\begin{table}[htb!]
    \centering
    \caption{Severity categories used to assign risk under MIL-STD-882E \cite{MILSTD882E}}
    \begin{tabular}{|p{2cm}|p{1.2cm}|p{11cm}|}\hline
         Description & Severity & Mishap Result Criteria \\\hline
         Catastrophic & 1 & Death, permanent total disability, irreversible significant environmental impact, or monetary loss $\geq$ \$10M. \\\hline
         Critical & 2 & Permanent partial disability, injuries or occupational illness that may result in hospitalization of at least three personnel, reversible significant environmental impact, or monetary $\geq$\$1M and $<$ \$10M. \\\hline
         Marginal & 3&  Injury or occupational illness resulting in one or more lost work day(s), reversible moderate environmental impact, or monetary loss $\geq$ \$100K and $<$ \$1M. \\\hline
         Negligible & 4 & Injury or occupational illness not resulting in a lost work day, minimal environmental impact, or monetary $<$ \$100K.\\\hline
    \end{tabular}
    \label{tab:severity}
\end{table}

\section{Trust}
Within the trust in automation community, trust is defined as ``the attitude that an agent will help achieve an individual’s goals in a situation characterized by uncertainty and vulnerability" \cite{lee2004trust}. Trust is inherently an evaluation of whether the benefits outweigh the risk. In teaming scenarios, trust may occur between humans, between AIs or between humans and an AI. One of the biggest challenges is ensuring that trust between two agents (human or AI) is calibrated. \textit{Overtrust} may result in relying on an agent in a scenario they are ill-suited for. \textit{Undertrust} may result in lower system performance when the agent is not used in a situation they are better suited to handle. Each agent, whether human or automation must have a good representation of the strengths and weaknesses of the partner, which may be gained through experience.

\subsection{Trust Dynamics}
Trust is a dynamic rather than a static process. When it comes to humans and technology, the general default assumption is that the automation is perfect, initially resulting in overtrust. However, any mistake by the automation quickly erodes that trust \cite{madhavan2007similarities}. Additionally, high false alarm rates or nuisance alerts or activations of AI can result in the system being turned off \cite{sorkin1988people,ho2017trust}. In one study of pilot trust in an automatic ground collision avoidance system, a pilot indicated that one mistake by the automation would likely cause pilots to turn the system off \cite{ho2017trust}.

\subsection{Transparency}
To facilitate trust between human-human, human-AI, and AI-AI systems, each agent needs some model of the cooperating agent. Transparency could include an explanation of the agent's rationale, facilitate understanding of the agent's behavior, and highlight shared understanding of interaction \cite{chen2020guest,mercado2016intelligent}. As operators become familiar with a system and see its operation is consistent with how its function was explained to them, they gain positive impressions \cite{lyons2016trust}.

\subsection{Human-Automation Trust Variability}
Trust in automation varies from human to human. One trust categorization is situational, learned, or dispositional trust \cite{marsh2003role}.
\textit{Situational} trust describes an individual's level of trust for a specific context. For example, an operator may rely on automation more when overwhelmed with a high workload \cite{biros2004influence}.
\textit{Learned} trust is based on experiences with the specific automated system \cite{marsh2003role}, and changes over time \cite{hoff2015trust}.
\textit{Dispositional} trust is when an individual's propensity to trust automation is driven by culture or personality. For example, studies have found that citizens of Mexico are more likely to trust automated decision aids than citizens of the United States \cite{huerta2012framing}, older adults are more likely to have better calibrated trust of decision aids \cite{ho2005age, sanchez2004reliability} and are less likely to be swayed by a picture of a human expert in the automation interface \cite{pak2012decision}, women and men respond different to flattery and communication styles from automation \cite{lee2008flattery,nomura2008prediction,tung2011influence}. Trust can also be influenced by different personality categorizations. People are more likely to trust if they are extroverted \cite{merritt2008not} or intuitive \cite{mcbride2012impact}, and less likely if they have neurotic \cite{szalma2011individual} or sensing \cite{mcbride2012impact} personalities.


\subsection{Anthropomorphizing Autonomy}
Trust may also be influenced by a tendency to anthropomorphize AI and robots. While science fiction has long been concerned with anthropomorphizing robots \cite{robot}, real world examples of humans anthropomorphizing robots grew in the early 2000s. The first documented case of this is with iRobot's Roomba platform, a robot vacuum which some users gave positive traits and characteristics due to its function and proximity to users \cite{roomba}. Not only did users want to personalize their unique Roomba, operators assigned Roombas specific names and considered them similar to house pets, and became more willing to tolerate more extraneous behavior, compared to other cleaning robots \cite{roomba} due to their nonthreatening nature and positive contribution to the household. The Army Research Laboratory has expanded this concept further, finding similarities in Human-Autonomy Teaming and Human-Animal Trust \cite{humananimal}. If the robot interacted with humans in a friendly way on occasion, if the errors it made were not severe, if the operator underwent training with the robot, if it was "cute" in either behavior or design, and if the environment the human and robot operated in was non-threatening, operators were more likely to accept unpredictable behavior \cite{humananimal}. 

\subsection{Stakeholder Relationships}
Multiple studies have identified that mutual understanding and respect between stakeholders of a technology impact trust. Stakeholders may include researchers, regulators, program managers, design engineers, system acquirer, insurers, commander or supervisor, operator, patron, the community, and in some cases the science team and science community \cite{ho2017trust,lacher2014autonomy}. Each group features risks such as personal safety, property and finances, mission effectiveness, reputation, job security, and public trust. During the design process, researchers, program mangers, and design engineers are accountable for designing the system to perform well, are vulnerable to customer backlash including loss of reputation and future resources, and expect the system to be used as intended. Regulators and acquirers are responsible for accomplishing mission goals, vulnerable to a loss of required resources to complete design and evaluation, and expect the design to meet operator needs. Operators, patrons, the community, and in some cases, the scientific community, are accountable for using tools as intended to successfully complete the mission, vulnerable to failure, which could include loss of priceless science data, property, finances, reputation or life, and expect the AI system to perform acceptably and free from errors or biases that could call analysis of mission data into question \cite{ho2017trust,mandrake2022space}.

\subsection{Cross-domain Examples of How Trust in Automation Impacts Safety}

A critical section of designing safety with autonomous systems that cannot be ignored is the impact of trust in interaction of the algorithm with the human. In particular, automation complacency (or alternatively, appropriate reliance) and poor transparency of the system operation to human users have contributed to major accidents across safety critical industries. 

\subsubsection{Automation Complacency}
The NTSB Highway Accident Report on the "Collision Between Vehicle Controlled by Developmental Automated Driving System and Pedestrian" \cite{ubercrashntsb}, which was a landmark for being the first ever investigation of a human death by self-driving vehicle, discovered one of the major factors leading to the crash was \textit{automation complacency}. Automation complacency is first described by Parasuraman \cite{paracomplacencybias}, and is a form of overtrust under the Lee and See category of appropriate reliance \cite{leeandsee}. \textit{Appropriate reliance} is related to \textit{calibrated trust}. When a user overtrusts an AI or automation solution, they can become complacent and use the AI in a context that it was not designed for. In cases where a human undertrusts AI, it may lead to a situation where the human stops using the AI in a context where it is better suited. Both over and undertrust can contribute to safety violations. The NTSB concluded that autonomous vehicles had high levels of automation complacency, and so began recommending all automated vehicle manufacturers monitor their drivers for attentiveness \cite{ubercrashntsb}. In response, car manufacturers, notably Tesla, Ford and GM, began to implement torque sensors, which required a human to hold the steering wheel, and are testing driver-facing cameras to detect if drivers become inattentive \cite{drivermonitoringtsla}. The debate over whether a system should monitor its users comes down to the culture of the operator and the kind of trust between the operator, which should be explored in future research.

\subsubsection{Appropriate Transparency}

Designing appropriate transparency that communicates system information and alarms in human-autonomy teams can be challenging. On March 28th, 1979, at 4:00AM, the Three Mile Island nuclear reactor began a series of events which led to America's worst nuclear meltdown. The US Presidential Commission on the Accident at 3 Mile Island found 3 major factors that caused the meltdown, one of which was not a design flaw in the reactor, but in the operator console \cite{tmireport}. The Three Mile Island plant was one of the most complex computer controlled systems of its time, but the investigation found that the system did not give clear warnings of incidents, or explain what the system was doing in response \cite{tmireport}. Another example of the need for transparency came from the unfortunate set of crashes in 2019 related to the Boeing 737 Max 8. The NTSB noted in their Safety Recommendation Report \cite{boeingcrashntsb} that one of the primary failures leading to the crashes was improper training of flight crew and withholding of information to flight personnel \cite{boeingcrashntsb}. Clear and appropriate transparency, combined with good operator training, is a requirement for human-autonomy teaming in automated and complex situations.

\section{Ethics}
In this section, we summarize the ethical viewpoints of philosophy, research/engineering, and law on the use of AI through published documentation and literature. We expand on the primary ethical challenges researchers and engineers face in development, deployment, and use of machine learning. The authors of this manuscript argue that ethical \textit{use} of AI from design through operations is more appropriate and achievable than developing AI that has "ethics." This section explores some of the prominent AI ethics concepts and historical examples, but is intended as a primer rather than a comprehensive review.

\subsection{Philosophical Machine Learning Ethics}

Ethics in philosophy concerns the morality of decisions, of right and wrong, good and evil. For the last decade, philosophers have become concerned with the rise of AI; in particular, how to apply concepts of right and wrong to AI. In general, philosophers tackling ethics subscribe to one of the three frameworks \cite{ethicsHAIT}, shown in Table \ref{tab:ethicframeworks}. These philosophical considerations and help build the foundation of trust. Consider that a person in need should be helped. A deontologist might argue that helping aligns with the golden rule (e.g. "do unto others as you would have them do unto you"), a virtue ethicist might argue that helping is charitable or benevolent, and a consequentialist might argue that helping maximizes well-being \cite{Hursthouse2022Virtue}. 

\begin{table}[hbt!]
\caption{Types of Ethics in Philosophy \cite{ethicsHAIT} \label{tab:ethicframeworks}}
\centering
\begin{tabular}{|p{0.2\linewidth} | p{0.6\linewidth}|}
\hline
Ethical Framework& Description \\\hline
Deontology& Right and wrong are set by a predetermined set of rules governing good and bad intent in conformance with moral norms.\\
Virtue Ethics& Ethical decisions are driven by the moral character of a person or group. \\
Consequentialism & Whether an action is ethically right depends only on actual consequences.\\
\hline
\end{tabular}
\end{table}

\subsubsection{The Golem and Trust: Believing Humans Can Anticipate all Scenarios is Moral Hubris}
The first story of ethical use of AI is arguably the 16th century Jewish myth, ``The Golem of Prague" \cite{Spinoza}. The most common version is that a golem was created to defend the city of Prague, but the golem is sealed for various reasons. The most relevant reason to AI is that the golem misinterprets it's programming and performs an action which is unexpected \cite{praguegolemhawaii}. This breaking of trust between the golem and its users is communicated as a moral of hubris \cite{golem-umich}, but also demonstrates how ethical behavior factors into trust. 

\subsubsection{The 21st Century Trolley Problem: Not all Cultures Share the Same Ethics}
The trolley problem is a well known moral psychology question which was first introduced in 1967 asking the reader to suppose a man "is the driver of a runaway tram which he can only steer from one narrow track onto another; five men are working on one track and one man on the other; anyone on the track he enters is bound to be killed" \cite{foot1967problem}. Which track should he steer the tram (trolley) onto? The trolley problem was later formalized and expanded on with several variations \cite{thomson1984trolley}. While initially just a starter for ethics debates, modern scientists use data driven psychology in order to understand the trolley dilemma, specifically by polling large groups of people on the situation in order to break down morality by location, gender, income gap, ethnicity, and other factors \cite{michigantrolley}. MIT has put together their own version, asking users what kind of decision an autonomous car should make given varying factors \cite{mitmoral}. MIT was able to break down how the public would see an autonomous car crash by ethnicity, gender, religion, and affiliation in their paper \textit{The MIT Moral Machine} \cite{mitmoral}. MIT and other ethics researchers have shown that different societal groups are likely to have different ethical interpretations.

\subsubsection{Asimov and the \rev{Three Laws of Robotics}}
The most well known ethical boundary in robotics and automation was created by Issac Asimov in 1942, known as the \rev{three} laws of robotics which feature primarily in \rev{two} books: \textit{I, Robot} \cite{robot} and \textit{Foundations} \cite{foundation}. In his stories, Asimov proposes the following \rev{three} laws be programmed as the foundational starting seed for robots to behave ethically \cite{foundation}. These rules are\rev{:}

\vspace{10pt} \begin{enumerate}
  \item A robot may not injure a human being or, through inaction, allow a human being to come to harm. 
  \item A robot must obey the orders given it by human beings, except where such orders would conflict with the First Law. 
  \item A robot must protect its own existence as long as such protection does not conflict with the First or Second Law.
\end{enumerate} \vspace{10pt}

\rev{ Much of \textit{I, Robot} \cite{robot}, explores how modifying rule strength, or reordering rules leads to unintended consequences.} The \rev{three} laws are just as much of a societal expectation as they are a programmed rule, as Asimov mentions in \textit{Lost Little Robots} \cite{robot} and \textit{The Caves of Steel} \cite{robot}. He explores this in greater detail in \textit{Runaround} \cite{robot}, and then how those rules are manipulated by differing groups with different ethics in \textit{The Naked Sun} \cite{robot}\cite{foundation}. 

\subsubsection{Machine Learning Ethics in Philosophy Summary}
\rev{T}he Golem, the MIT Moral Machine, and Asimov's writings demonstrate 
\rev{a modern philosophical approach to ethics for autonomous decision making agents.} What 
\rev{one} individual considers 
a rational or ethical decision can be up for interpretation. However, as a whole, societies and organizations have differences in what they define as moral and their ethical boundaries. Therefore, any algorithm attempting to resolve ethical dilemmas 
should share the ethical boundaries of its users\rev{.} 
\rev{Designing for the ethics of users, rather than designers, may} 
be considered controversial in cases where algorithms are applied outside of their country of origin, but the alternative is a lack of trust between the user and the algorithm. 
If practical, users should be polled during the development of algorithms to factor in their ethical considerations before wide scale deployment.

\subsection{Machine Learning Ethics in Research and Engineering}
While philosophy considers the ethical and moral boundaries of AI, ethics in research and engineering \rev{place more emphasis on ethical usage. }
Most engineering literature define ethics as rules of right and wrong, following the philosophical principles of deontology \cite{asmeethics}. In this subsection we \rev{introduce the challenge of ethical design and usage of AI systems.} 

\subsubsection{Ethical Behavior in Machine Learning}

\rev{D}esigning for ethical use of algorithms during development and deployment \rev{is an active area of research}. The primary focus as of publishing has been on \rev{three} subjects: 1) whether algorithms can be programmed with ethical behavior, 2) \rev{how to remove }
bias in algorithm data, and 3) how organizational ethics, particularly around privacy, affect the human-AI team decisions. 

\textit{Ethics in Human-AI Teaming} \cite{ethicsHAIT} suggests that ML algorithms can and should be programmed with an ethics decision algorithm. However, others argue that since it is not really possible to code those principles to ML \cite{forbesethics}. David M. Tate of the Institute for Defense Analysis \cite{idapaper} also postulates that ethics is more about trust and transparency. In all cases, calibrated trust is required between the human and the ML to not violate the human's ethical boundaries \cite{ho2017trust}. 

\begin{table}[hbt!]
\caption{Types of Bias in AI \cite{nistbias} \label{tab:table2}}
\centering
\begin{tabular}{|p{0.14\linewidth} | p{0.8\linewidth}|}
\hline
Type of Bias& Description \\\hline
Systemic Bias& Systemic bias comes from information sourced from history, society, or institutions. This mostly manifests in the data that arrives from these sources and the bias this data was collected with. \\
Human Bias& Human bias comes from the user and the biases the user has towards their interaction, as the bias of the user or organization will manifest in the AI algorithm.\\
Statistical Bias& Statistical bias comes from the development of the AI, particularly in the use of the system, the sampling and filtering of the algorithm, and the AI model selection.\\
\hline
\end{tabular}
\end{table}

\begin{figure} [hbt!]
\centering
\includegraphics[width=.5\textwidth]{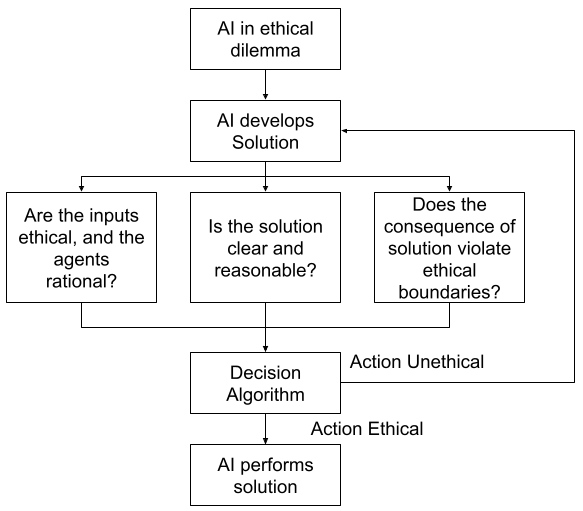}
\caption{\label{chart:ADCModel}Flow chart of modified Agent, Deed, Consequence Model}
\end{figure}

Another important problem in ethical design of ML models is removing bias. 
The National Institute for Standards and Technology (NIST) has put together a table of possible bias, \cite{nistbias} shown in Table \ref{tab:table2}. When not managed, bias can be amplified into bad behavior in ML \cite{ricespeechethics} \cite{neartermAIethicalmatrix}. To mitigate \rev{bad behavior, one approach is to} 
use feedback generated by the user and stakeholders to evaluate ethical behavior \cite{googlelamda} \cite{neartermAIethicalmatrix}, while \rev{another approach \cite{ethicsHAIT} is to build }
in a series of ethical safety checks during the training process, as demonstrated in Figure \ref{chart:ADCModel}.

\subsubsection{Responsible Use of ML}
As ML and AI become more prolific, the debate over using AI has begun to shift away from whether AI can be ethical in general and into whether AI is used responsibly in particular applications. For example, there are ethical concerns surrounding the  the development and deployment of Chatbots \cite{googlelamda}. 
Unfortunately, examples of chatbots "behaving unethically" have become more prolific due to human bias \cite{microsofttay}, or unethical applications \cite{4chanai}. Facial recognition is another controversial application, considering it's potential use cases and biases \cite{gendershades}. One significant concern, however, is focused on privacy and ethics of machines in defense and combat roles \cite{axonresignation}. 
\rev{The U.S. Department of Defense has defined tactical autonomy as a system acting with delegated and bounded authority given to it by humans \cite{autohorizons}.} 

\subsection{Differences in Ethical Perceptions Across Cultures}
Ethical use of AI is not universally defined, and two particularly different approaches depend on Eastern or Western philosophy. While there are many differences, this section will discuss differences in cultural philosophies, attitudes towards AI, trust in institutions and concerns over surveillance and privacy. According to a 2020 Pew Research Study, the public of Eastern countries have a more favorable view of AI (e.g. Singapore (72\%), South Korea (69\%), India (67\%), Taiwan (66\%) and Japan (65\%)), while Western countries are more split with lower favorable views (e.g. US (47\%), UK (46\%), Germany (47\%), and France (37\%)) \cite{johnson2020people}. The authors acknowledge that this is a complex issue, and that the brevity of this section prohibits comprehensive analysis. Additionally, generalizations of Western cultural views does not capture the nuanced differences in privacy protection policies, such as the EU's General Data Protection Regulation (GDPR) compared to fragmented privacy protection laws in the US \cite{almeida2022ethics}. 


\subsubsection{Example 1: Chinese AI Investment Strategy}
When Google Deepmind's AlphaGo defeated a professional Go player in 2016 \cite{silver2017mastering}, it was watched live by more than 280 million people in China and described by some as a "Sputnik moment," triggering new national strategies around AI, while the event was not nearly as significant to Western cultures \cite{roberts2021chinese}. The 2017 “New Generation Artificial Intelligence Development Plan” Strategy \cite{ChinaAIdevplan} aimed to make China the AI leader and with a trillion yuan industry by 2030, has an incentives list of technologies, described broad usage in sectors ranging from defense to social welfare, and became a driving force in ethical norms for AI in China \cite{roberts2021chinese}. 
Concern over "moral decline" in China outweighs the concerns for privacy culturally, providing the impetus to use AI for moral governance through the Social Credit System, which regulates citizen behavior with regard to facets such as finances, taxes, academic dishonesty, and behavior on public transit \cite{roberts2021chinese}. 
Chinese companies have cooperated more on creating and applying surveillance technology due to the passing of the National Intelligence Law of the People's Republic \cite{NILChina}, which broadly mandates all Chinese companies, citizens, and companies with assets in China to collaborate with a Chinese intelligence agency \cite{rethinkingchina}.


\subsubsection{Example 2: Westerns Concerns over Ethical use of AI}
A large cultural difference in the West is trust in institutions and their ability to ethically use AI. For example, according to a 2018 Pew Research study, the percent of US adults that believe many institutions behave unethically all or most of the time is 
high (e.g. members of congress (81\%), journalists (66\%), leaders of technology companies (77\%), religious leaders (69\%), police officers (61\%), military leaders (50\%), and local elected officials (66\%)) \cite{gecewicz2019americans}. Americans (82\%) also believe that robots and/or AI should be carefully managed (similar to EU survey results), and that the highest importance issues were \cite{zhang2019artificial}:
\begin{itemize}
    \item preventing AI-assisted surveillance from violating privacy and civil liberties,
    \item preventing AI from being used to spread fake and harmful content online,
    \item preventing AI cyberattacks against governments, companies, organizations, and individuals, and
    \item protecting data privacy.
\end{itemize}
These concerns are expanded upon in the October 2023 "Executive Order on the Safe, Secure, and Trustworthy Development and Use of Artificial Intelligence," which places particular attention on mitigating AI risks in areas such as biotechnology, cybersecurity, critical infrastructure, national security, the workforce, civil rights, consumer protections, and fair competition in AI development across large and small businesses \cite{Biden2023Executive}.

It's worth noting that western culture is concerned with the collaboration between defense and civilian technology. American attitudes towards AI, in particular autonomous weapons, are influenced, at least in part, by the films that depict armed AI systems \cite{young2018does}.
An example of this concern in the US is the public and corporate reaction to the DoD when it asked Google to work on Project Maven in 2017. In order to analyze the massive amount of video and other data generated from the growth in unmanned aerial vehicles \cite{usmamaven}, US Special Operations Command created a special joint task force that opened to industry to help solve the problem. Teams within Google, Microsoft, and Amazon Web Services were contracted to support the project \cite{forbesmaven}. However, major protests occurred within each company over worker concerns that their work may contribute to automated warfare \cite{msftmaven}. Protests drove Google to withdraw from the contract \cite{mavengoogle}. 

Germany and other EU nations faced similar worries over military adoption of AI platforms. In 2020 a German parliamentary spokesperson resigned alongside protests against the adoption of armed UAVs in the Bundeswehr \cite{germanUAVs}. Debate over whether the German military should use armed autonomous vehicles polarize national political debate in the EU, and show the level of public concerns over military use of autonomous and AI platforms \cite{germanydebate}.



While the concerns of philosophy and ethical design/use are to be considered, an organization's ethical framework and frameworks from law, affect the ethical use case and responsibility of operating ML agents in application. 
%
The United States Department of Defense's Innovation Board released a series of 5 guidelines of ethical principles that were officially adopted in 2020 \cite{dodethicalprinciples} and align with civilian organizations, 
as shown in Table \ref{tab:govethics}.

\section{Conclusion}
This publication serves as a primer on considerations for safe, trusted, and ethical use of AI in aerospace control systems. While these concepts are closely related, and presence of each strengthens the others, there are some nuanced differences in considerations. A brief introduction was provided on AI techniques such as reinforcement learning, with examples of how it is being applied in aerospace research. There was also discussion of what it means for humans in a human-AI team to be in, on, or out-of-the-loop. Next, specific considerations and references for safety, trust and ethics were presented. For safety, existing aerospace standards, design practices, and relationship to risk was presented. For trust, a discussion of the dynamic nature of trust, need for AI transparency, variability in automation trust between humans for a variety of reasons, the risks associated with anthropomorphizing AI, trust built between AI stakeholders, and examples of lessons learned across domains was presented. For ethics, different ethical frameworks and famous examples of ethical principles guiding use of AI were introduced, as well as types of bias, methods of analyzing consequence, cultural differences, and principles guiding capability development. AI and its applications to domains such as aerospace are evolving rapidly. Public access to generative AI tools like ChatGPT, DALL-E, Midjourney, Stable Diffusion, Google's Bard, and others in the last couple of years has led to a corresponding growth in public interest and discourse in the use of AI. Safe, ethical, and trusted use of AI in aerospace will likely also rapidly evolve with the rest of the community as AI technologies continue to mature.

\section*{Acknowledgments}
The views expressed are those of the authors and do not reflect the official guidance or position of the United States Government, the Department of Defense, or the United States Air Force.

\begin{table}[hbt!]
\caption{\label{tab:govethics} DoD and NASA Principles that Guide AI Capabilities  \cite{dodethicalprinciples,nasaaiethics}}
\centering
\begin{tabular}{|p{0.45\linewidth} | p{0.45\linewidth}|}\hline
\textbf{Department of Defense Innovation Board}& \textbf{NASA Guidelines}\\\hline

\textit{Responsible}: DoD personnel will exercise appropriate levels of judgment and care, while remaining responsible for the development, deployment, and use of AI capabilities. 

& \textit{Accountable}: Organizations and individuals must be accountable for the systems they
create, and organizations must implement AI governance structures to provide oversight. AI
developers should consider potential misuse or misinterpretation of AI-derived results
(intentional or otherwise) and take steps to mitigate negative impact.\\\hline

\textit{Equitable}: The Department will take deliberate steps to minimize unintended bias in AI capabilities. 
& \textit{Fair}: AI systems must include considerations regarding how to treat people, including
refining solutions to mitigate discrimination and bias, preventing covert manipulation, and
supporting diversity and inclusion.\\\hline

\textit{Traceable}: The Department’s AI capabilities will be developed and deployed such that relevant personnel possess an appropriate understanding of the technology, development processes, and operational methods applicable to AI capabilities, including with transparent and auditable methodologies, data sources, and design procedure and documentation. 
& \textit{Explainable and Transparent:} Solutions must clearly state if, when, and how an AI system
is involved, and AI logic and decisions must be explainable. AI solutions must protect intellectual
property and include risk management in their construction and use. AI systems must be
documented.\\\hline

\textit{Governable: }The Department’s AI capabilities will have explicit, well-defined uses, and the safety, security, and effectiveness of such capabilities will be subject to testing and assurance within those defined uses across their entire life-cycles. 
& \textit{Human-Centric:} AI systems must obey human legal systems and must provide benefits to society. At the current state of AI, humans must remain in charge, though future advancements may cause reconsideration of this requirement.
 \\\hline

\textit{Reliable:} The Department’s AI capabilities will have explicit, well-defined uses, and the safety, security, and effectiveness of such capabilities will be subject to testing and assurance within those defined uses across their entire life-cycles. 
& \textit{Secure and Safe:} AI systems must respect privacy and do no harm. Humans must monitor
and guide machine learning processes. AI system risk tradeoffs must be considered when
determining benefit of use.\\\hline

& \textit{Scientifically and Technically Robust:} AI systems must adhere to the scientific method
NASA applies to all problems, be informed by scientific theory and data, robustly tested in
implementation, well-documented, and peer reviewed in the scientific community.\\\hline
\end{tabular}
\end{table}



\bibliography{references}

\end{document}